# Chiral two-dimensional periodic blocky materials with elastic interfaces: auxetic and acoustic properties

Andrea Bacigalupo and Luigi Gambarotta[*]

Department of Civil, Chemical and Environmental Engineering, University of Genova, Italy

**Abstract**

Two novel chiral block lattice topologies are here conceived having interesting auxetic and acoustic behavior. The architectured chiral material is made up of a periodic repetition of square or hexagonal rigid and heavy blocks connected by linear elastic interfaces, whose chirality results from an equal rotation of the blocks with respect to the line connecting their centroids. The governing equation of the Lagrangian model is derived and a hermitian eigenproblem is formulated to obtain the frequency band structure. An equivalent micropolar continuum is analytically derived through a standard continualization approach in agreement with the procedure proposed by Bacigalupo and Gambarotta (2017) from which an approximation of the frequency spectrum is derived. Moreover, the overall elastic moduli of the equivalent Cauchy continuum are obtained in closed form via a proper condensation procedure. The parametric analysis involving the overall elastic moduli of the Cauchy equivalent continuum model and the frequency band structure is carried out to catch the influence of the chirality angle and of the ratio between the tangential and normal stiffness of the interface. Finally, it is shown how chirality and interface stiffness may affect strong auxeticity and how the equivalent micropolar model provides dispersion curves in excellent agreement with the current ones for a wide range of the wave vector magnitude.

---

[*] Corresponding Author, luigi.gambarotta@unige.it

# 1. Introduction

Exotic mechanical properties, such as auxetic behavior and tunable frequency band structure, are peculiar of innovative architected materials resulting from a complex interplay of topology, material distribution and constituent material behavior which nowadays may be obtained through the new techniques of addictive manufacturing. Artificial materials with properly arranged periodic microstructure may exhibit auxetic behavior, i.e. negative Poisson ratio, have been designed and manufactured (see Saxena *et al.*,2016, Lakes 2017, Ren *et al.*,2018, for reference). Moreover, the topologies thus obtained may present interesting acoustic responses, to be exploited in the realization of acoustic meta-filters and wave propagation controllers (see for reference Cummer *et al.*,2016, Deymier, 2013, Zangeneh-Nejad and Fleury, 2019). A wide class of such materials is based on the lattice microstructure composed by elastic ligaments properly arranged in architected periodic cells (see for instance the recent contributions by Chen and Fu, 2017, Lu *et al.*, 2017, Li *et al.*, 2019, Vila *et al.*, 2017, Phani and Hussein,2017). Since the seminal paper of Prall and Lakes (1997) the chiral topology of the periodic cell in lattices has shown to provide an interesting auxetic behavior (see Liu *et al.*, 2012, Bacigalupo and Gambarotta, 2014, Ha *et al.*, 2006, Wu *et al.*,2019, Attard *et al.*, 2020, among the others).

Periodic microstructures having a rigid phase with dominant volumetric fraction connected by elastic devices with vanishing volume fraction have been considered and analysed by several Authors. Grima *et al.* (2004, 2005) proposed a novel microstructure consisting in rigid polygonal blocks connected at their vertices by elastic hinges arranged according to chiral configurations to achieve a strong auxetic behavior, with Poisson ratio equal to -1. Developments of these polygonal rotating mechanisms have been proposed by Gatt *et al.* (2015) and Dudek *et al.* (2020) among the others. Cellular flexible metamaterials have been proposed by Mizzi *et al.* (2018) and Liang and Crosby (2020), who conceived two-dimensional plate-chain models of the periodic cell, namely chains of elastic or rigid polygonal plates connected at the vertices by flexible beams, to obtain strong auxeticity. Tridimensional rotating mechanisms have been recently proposed by Tanaka *et al.* (2020) in which the unit cell is a stellated octahedron made up of tetrahedral units connected at the vertices with linear springs. Other auxetic materials have been proposed which are based on lattice structures defined by patterns of slits that follow chiral configurations. In these materials the slits pattern creates deformation mechanisms leading to negative Poisson ratios (see Zhang *et al.*,2018). Bilski *et al.* (2016) proposed assemblies of periodic rigid particles with variable shape to obtain a material having auxetic behavior based on two-dimensional systems of hard discs with parallel layers of hard cyclic hexamers. An interesting analysis concerning materials based on the two-phase infilled re-entrant honeycomb microstructure has been carried out by Peng *et al.*, 2020, where the auxetic behavior is found to be strongly dependent on the mismatch of the Young modulus between the two components. When the re-entrant lattice is stiffer than the filler the auxeticity stems from the conventional re-entrant mechanism, conversely if the re-entrant lattice is softer than the filler the auxeticity results from the relative sliding in the filler.

Bacigalupo and Gambarotta (2017) analyzed the macroscopic and acoustic properties of periodic assemblies of rigid rhombic and hexagonal blocks connected by elastic interfaces. In the case of hexagonal tiling it has been shown that the auxetic behavior is attained if the axial stiffness of the interfaces is lesser than the tangential one. A circumstance that can only be realized through interfaces such as that proposed by Shufrin *et al.* (2012) who studied the elastic properties of assemblies of rigid hexagons connected to each other by suitable curvilinear beams. In this framework, two novel chiral block lattice topologies are here proposed having interesting auxetic and acoustic behavior (see Figure 1). The periodic material is obtained as a regular repetition of square or hexagonal rigid and heavy blocks connected by (micro-structured) linear elastic interfaces. The blocks are rotated in the periodic pattern to obtain chiral configurations. The governing equation of the Lagrangian model is derived and the frequency band structure is obtained. The constitutive tensors of the equivalent micropolar continuum are derived through a standard continualization approach in agreement with the procedure proposed by Bacigalupo and Gambarotta (2017) and an approximation of the frequency spectrum is analytically derived. The overall elastic moduli of the equivalent Cauchy continuum are derived via a proper condensation procedure of the micropolar constitutive equation according to Bacigalupo and Gambarotta (2016). The parametric analysis here developed show the influence of chirality and interface elastic stiffness on the global Poisson ratio, on the elastic stiffness and on the frequency spectrum of the material. Finally, it is shown how chirality and interface stiffness may affect strong auxeticity and how the equivalent micropolar model provides dispersion curves in excellent agreement with the current ones for a wide excursion of the wave vector magnitude.

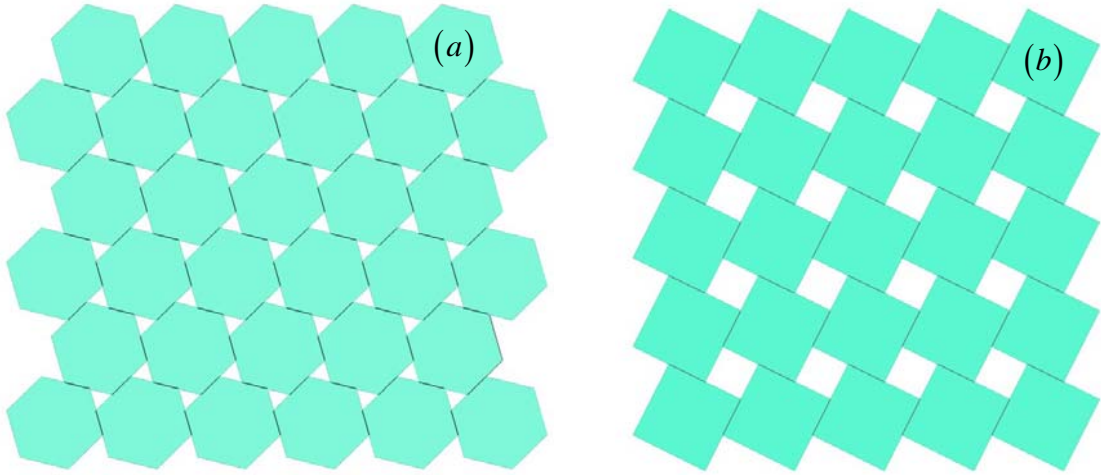

Figure 1: Chiral patterns of rigid blocks connected with elastic interfaces: (a) hexagonal ($n=6$) and (b) square ($n=4$).

## 2. The chiral periodic blocky material: the discrete and the micropolar model

Let consider a two-dimensional periodic assemblage of rigid regular polygons as shown in the Figure 1 connected along the contact line through a linear elastic interface. Here assemblages of hexagonal (coordination number $n=6$) and square blocks ($n=4$) are considered, which are characterized by a chiral geometry measured by the rotation $\beta$ of each block with respect to the segment connecting its centre with the centre of the surrounding block.

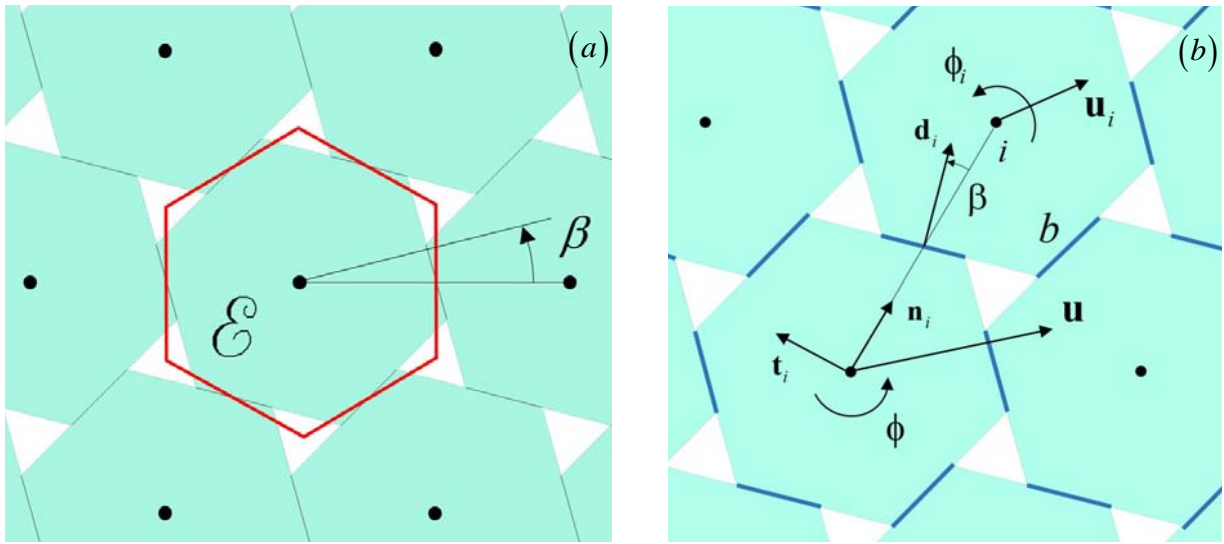

Figure 2. Chiral hexagonal blocky assemblage: (a) periodic cell; (b) dofs of the discrete model.

Due to the pattern periodicity a reference block is considered and a periodic cell $\mathcal{E}$ is identified (see Figure 2a). The reference block has area $A_b$ and is equipped with mass $M$, mass density $\rho_b = M/A_b$ and moment of inertia $J = Mr_b^2$ (a unit thickness of the blocks is here assumed). Its motion is described through the displacement $\mathbf{u}(t)$ and rotation $\phi(t)$ of the block center, respectively (see Figure 2b), while the motion of the connected $i$-th block is expressed by the translation $\mathbf{u}_i(t)$ and the rotation $\phi_i(t)$ ($i=1,n$). The location of the centre of the $i$-th block with respect to the reference block is given by vector $\mathbf{x}_i = l\mathbf{n}_i$, $l$ being the distance between adjacent blocks and $\mathbf{n}_i$ the unit vector identifying the connection line (see Figure 2b). Moreover, the $\mathbf{t}_i = \mathbf{e}_3 \times \mathbf{n}_i$ unit vector normal to $\mathbf{n}_i$ and the $\mathbf{d}_i$ unit vector normal to the interface are introduced. The interfaces have size $b$ with vanishing thickness and normal $k_n$ and tangential $k_t$ stiffness per unit length, respectively. The relative displacement $\Delta \mathbf{s}_i$ and relative rotation $\Delta \varphi_i$ at the midpoint of the interface connecting the reference block and the adjacent $i$-th block are considered, the former being the superposition of the normal relative displacement $\Delta \mathbf{s}_{ni} = \mathbf{P}_i^{\parallel} \Delta \mathbf{s}_i$ and the tangential relative displacement $\Delta \mathbf{s}_{ti} = \mathbf{P}_i^{\perp} \Delta \mathbf{s}_i$ with $\mathbf{P}_i^{\parallel} = \mathbf{d}_i \otimes \mathbf{d}_i$ and $\mathbf{P}_i^{\perp} = \mathbf{I} - \mathbf{d}_i \otimes \mathbf{d}_i$ projection tensors associated to vector $\mathbf{d}_i$. The elastic energy stored in the $i$-th interface may be written as

$$\pi_i = \frac{1}{2} K_n \Delta \mathbf{s}_{ni}^2 + \frac{1}{2} K_t \Delta \mathbf{s}_{ti}^2 + \frac{1}{2} K_\varphi \Delta \varphi_i^2, \tag{1}$$

being $K_n = k_n b$, $K_t = k_t b$ and $K_\varphi = k_n b^3/12$ the normal, tangential and rotational overall stiffness of the interface, respectively, while the kinetic energy of the reference block is

$$T(\dot{\mathbf{u}}, \dot{\phi}) = \frac{1}{2} M \dot{\mathbf{u}}^2 + \frac{1}{2} J \dot{\phi}^2. \tag{2}$$

Following the Lagrangian treatment proposed by Bacigalupo and Gambarotta, 2017, the equations of motion of the reference block take the form of three ODEs

$$\sum_{i=1}^{n} \left[ \mathbf{K}_i (\mathbf{u} - \mathbf{u}_i) + \frac{l}{2} \mathbf{K}_i \mathbf{t}_i (\phi + \phi_i) \right] + M_b \ddot{\mathbf{u}} = \mathbf{0}, \tag{3}$$

$$\sum_{i=1}^{n} \left[ \frac{l}{2} \mathbf{t}_i \cdot \mathbf{K}_i (\mathbf{u} - \mathbf{u}_i) + K_\varphi^i (\phi - \phi_i) + \frac{l^2}{4} \mathbf{t}_i \cdot \mathbf{K}_i \mathbf{t}_i (\phi + \phi_i) \right] + J \ddot{\phi} = 0. \tag{4}$$

In equation (3) and (4) the second order tensor is defined

$$\mathbf{K}_i = \bar{K}\, \mathbf{n}_i \otimes \mathbf{n}_i + 2\tilde{K}\, \text{sym}(\mathbf{t}_i \otimes \mathbf{n}_i) + \hat{K}\, \mathbf{t}_i \otimes \mathbf{t}_i, \tag{5}$$

being

$$\begin{aligned}
\bar{K} &= K_n \cos^2 \beta + K_t \sin^2 \beta, \\
\tilde{K} &= \frac{(K_n - K_t)}{2} \sin 2\beta, \\
\hat{K} &= K_n \sin^2 \beta + K_t \cos^2 \beta,
\end{aligned} \tag{6}$$

constitutive parameters depending on the chirality angle $\beta$.

The harmonic waves travelling along unit vector $\mathbf{e}$ are analysed by assuming the standard form of the displacement field $\mathbf{U} = \hat{\mathbf{U}} \exp[i(\mathbf{k} \cdot \mathbf{x} - \omega t)]$, being $\mathbf{k} = q\mathbf{e}$ the wave vector, $q$ and $\omega$ the wave number and the circular frequency, respectively, and $\hat{\mathbf{U}} = \{\hat{\mathbf{u}}^T \quad \hat{\phi}\}^T = \{\hat{u}_1 \quad \hat{u}_2 \quad \hat{\phi}\}^T$ the polarization vector. From this assumption and due to the center-symmetry of the microstructure the following eigen-problem governing the harmonic wave propagation is obtained

$$\left[\mathbf{C}_{Lag}(\mathbf{k}) - \omega^2 \mathbf{M}\right]\hat{\mathbf{U}} = \left\{\begin{bmatrix} \mathbf{A} & i\mathbf{b} \\ -i\mathbf{b}^T & C \end{bmatrix} - \omega^2 \begin{bmatrix} M\mathbf{I}_2 & \mathbf{0} \\ \mathbf{0} & J \end{bmatrix}\right\} \begin{Bmatrix} \hat{\mathbf{u}} \\ \hat{\phi} \end{Bmatrix} = \mathbf{0}, \tag{7}$$

where the hermitian matrix $\mathbf{C}_{Lag}(\mathbf{k})$ has the following submatrices

$$\begin{aligned}
\mathbf{A} &= \sum_{i=1}^{n} \left[1 - \cos(\mathbf{k} \cdot \mathbf{x}_i)\right] \mathbf{K}_i, \\
\mathbf{b} &= \sum_{i=1}^{n} \left[\frac{l}{2} \sin(\mathbf{k} \cdot \mathbf{x}_i) \mathbf{K}_i \mathbf{t}_i\right], \\
C &= \sum_{i=1}^{n} \left\{K_\varphi \left[1 - \cos(\mathbf{k} \cdot \mathbf{x}_i)\right] + \frac{l^2}{4} \hat{K} \left[1 + \cos(\mathbf{k} \cdot \mathbf{x}_i)\right]\right\}.
\end{aligned} \tag{8}$$

Three dispersive functions $\omega(\mathbf{k})$ are obtained as solutions of the eigen-problem (7), which define two acoustic branches and an optical branch in the Floquet-Bloch spectrum (Brillouin, 1953). For long waves limit $q \to 0$ it follows $\mathbf{A} = \mathbf{0}$, $\mathbf{b} = \mathbf{0}$, $C_0 = \lim_{q \to 0} C(q) = n\hat{K}l^2/2$ and from equation (7) two vanishing frequencies $\lim_{q\to 0}\omega_{ac1,2}(q)=0$ and a critical point $\lim_{q\to 0}\omega_{opt}(q)=\sqrt{C/J}=\sqrt{n\hat{K}l^2/2J}$ with vanishing group velocity are obtained.

The discrete chiral blocky material may be homogenized to obtain a micropolar continuum description and to provide synthetic constitutive data, such as the elasticity tensors and moduli. This goal may be achieved through a second order continualization of the nodal generalized displacements of the blocks based on a macro-displacement $\mathbf{v}(\mathbf{x},t)$ and a macro-rotation $\theta(\mathbf{x},t)$ continuum fields, respectively. The assumed down-scaling law is

$$\mathbf{u}_i(t) \cong \mathbf{v}(\mathbf{x},t) + l\, \mathbf{H}(\mathbf{x},t)\mathbf{n}_i + \frac{1}{2}l^2\, \nabla\mathbf{H}(\mathbf{x},t):(\mathbf{n}_i \otimes \mathbf{n}_i),$$
$$\phi_i(t) \cong \theta(\mathbf{x},t) + l\, \boldsymbol{\chi}(\mathbf{x},t)\cdot\mathbf{n}_i + \frac{1}{2}l^2\, \nabla\boldsymbol{\chi}(\mathbf{x},t):(\mathbf{n}_i \otimes \mathbf{n}_i),$$
(9)

being $\mathbf{H}=\nabla\mathbf{v}$, $\nabla\mathbf{H}$ the second gradient of the macro-displacement, $\boldsymbol{\chi}=\nabla\theta$ and $\nabla\boldsymbol{\chi}$ the first and second gradient of the rotation, respectively. According to Bacigalupo and Gambarotta (2017) the continualized equation of motion are those of the Cosserat continuum as follows

$$\nabla\bullet(\mathbb{E}\,\boldsymbol{\Gamma}) = \rho\ddot{\mathbf{v}},\qquad(10)$$

$$\nabla\bullet(\mathbf{E}\,\boldsymbol{\chi}) - \in_{3jh}(\mathbf{e}_j \otimes \mathbf{e}_h):(\mathbb{E}\,\boldsymbol{\Gamma}) = I\ddot{\theta},\quad j,h=1,2,\qquad(11)$$

where the strain tensor $\boldsymbol{\Gamma}=\mathbf{H}-\mathbf{W}(\theta)$ is introduced (see Nowacky, 1986, Eremeyev et al., 2013) and having defined $\rho = M/A_{cell} = \rho_b A_b/A_{cell}$ the mass density and $I = J/A_{cell} = \rho_b r_b^2 A_b/A_{cell} = \rho r_b^2$ the density of rotational inertia referred to the periodic cell $\mathbf{E}$ of area $A_{cell}$, respectively. Moreover, the fourth order elasticity tensor equipped with major symmetry

$$\mathbb{E} = \frac{l^2}{2A_{cell}} \sum_{i=1}^{n} \left[ \begin{array}{c} \bar{K}\,\mathbf{n}_i\otimes\mathbf{n}_i\otimes\mathbf{n}_i\otimes\mathbf{n}_i + \tilde{K}\,(\mathbf{t}_i\otimes\mathbf{n}_i\otimes\mathbf{n}_i\otimes\mathbf{n}_i + \mathbf{n}_i\otimes\mathbf{n}_i\otimes\mathbf{t}_i\otimes\mathbf{n}_i) + \\ + \tilde{K}\,\mathbf{t}_i\otimes\mathbf{n}_i\otimes\mathbf{t}_i\otimes\mathbf{n}_i \end{array} \right],\qquad(12)$$

and the symmetric second order elasticity tensor

$$\mathbf{E} = \frac{l^2}{2A_{cell}} \sum_{i=1}^{n} \left( K_\varphi - \frac{l^2}{4}\hat{K} \right) \mathbf{n}_i \otimes \mathbf{n}_i.\qquad(13)$$

are defined, that depend on the chirality angle $\beta$ through equation (6). Because the centro-symmetry of the periodic cell the uncoupled constitutive equations hold $\mathbf{T} = \mathbb{E}\,\mathbf{\Gamma}$ and $\mathbf{m} = \mathbb{E}\,\mathbf{\chi}$, $\mathbf{T}$ and $\mathbf{m}$ being the micropolar stress tensor and the couple-stress vector, respectively. According to this continuum formulation, the propagation of harmonic waves is described by assuming the macro-displacement field in the form $\mathbf{V} = \hat{\mathbf{V}} \exp[i(\mathbf{k}\cdot\mathbf{x} - \omega t)]$ with $\hat{\mathbf{V}} = \{\hat{\mathbf{v}}^T \quad \hat{\theta}\}^T = \{\hat{v}_1 \quad \hat{v}_2 \quad \hat{\theta}\}^T$. As usual an eigen-problem is obtained $[\mathbf{C}_{\text{hom}}(\mathbf{k}) - \omega^2 \mathbf{M}]\hat{\mathbf{V}} = \mathbf{0}$ having the same structure of problem (7), with the submatrices of the hermitian matrix $\mathbf{C}_{\text{hom}}(\mathbf{k})$ analogous to equation (7) and defined as follows:

$$\mathbf{A}_{\text{hom}} = \frac{l^2}{2A_{cell}} \sum_{i=1}^{n} [(\mathbf{n}_i \otimes \mathbf{n}_i) : (\mathbf{k} \otimes \mathbf{k})] \mathbf{K}_i,$$

$$\mathbf{b}_{\text{hom}} = \frac{l^2}{2A_{cell}} \sum_{i=1}^{n} [(\mathbf{k}\bullet\mathbf{n}_i) \mathbf{K}_i \mathbf{t}_i], \qquad (14)$$

$$C_{\text{hom}} = \frac{l^2}{2A_{cell}} \sum_{i=1}^{n} \left\{ \hat{K} + \left[ K_\varphi - \frac{l^2}{4} \hat{K} \right] [(\mathbf{n}_i \otimes \mathbf{n}_i) : (\mathbf{k} \otimes \mathbf{k})] \right\}.$$

It is worth to note that $\mathbf{C}_{Lag}(\mathbf{k}) = A_{cell}\,\mathbf{C}_{\text{hom}}(\mathbf{k}) + \mathcal{O}(q^3)$. As a consequence, the dispersion functions $\omega_j(\mathbf{k})$, $j=1,3$ solution of the eigen-problem provide an approximation of those obtained from the discrete model in the long-wavelength limit $q \to 0$.

### 3. Hexachiral blocky system

Let consider now the chiral blocky system composed of hexagons with apothem $a/2$, area $A_b = \sqrt{3}a^2/2$ as shown in Figure 3. The hexagons are rotated by the chirality angle $\beta$ ($0 \leq \beta < \pi/6$). The distance between the centres of blocks in contact is $l = a/\cos\beta$ and the elastic interface measures is $b = (\sqrt{3}/3 - \tan\beta)a$. The periodic cell is a hexagon having apothem $l/2$ and area $A_{cell} = \sqrt{3}l^2/2 = \sqrt{3}a^2/(2\cos^2\beta)$. The inertia terms are $\rho_b = M/A_b$, $\rho = \rho_b \cos^2\beta$, $J = 5\sqrt{3}\rho_b a^4/72$, $r_b^2 = 5a^2/36$, $I = 5\rho a^2/36$. The overall elastic parameters of the interfaces are $K_n = (\sqrt{3}/3 - \tan\beta)k_n a$, $K_t = (\sqrt{3}/3 - \tan\beta)k_t a$ and $K_\varphi = (\sqrt{3}/3 - \tan\beta)^3 k_n a^3/12$, respectively. The unit vectors identifying the blocks surrounding the reference one are independent on the chirality angle and are $\mathbf{n}_1 = -\mathbf{n}_4 = \mathbf{e}_1$, $\mathbf{t}_1 = -\mathbf{t}_1 = \mathbf{e}_2$, $\mathbf{n}_2 = -\mathbf{n}_5 = 1/2\,\mathbf{e}_1 + \sqrt{3}/2\,\mathbf{e}_2$, $\mathbf{t}_2 = -\mathbf{t}_5 = -\sqrt{3}/2\,\mathbf{e}_1 + 1/2\,\mathbf{e}_2$, $\mathbf{n}_3 = -\mathbf{n}_6 = -1/2\,\mathbf{e}_1 + \sqrt{3}/2\,\mathbf{e}_2$, $\mathbf{t}_3 = -\mathbf{t}_6 = -\sqrt{3}/2\,\mathbf{e}_1 - 1/2\,\mathbf{e}_2$. The unit vectors normal to the interfaces are $\mathbf{d}_i = \mathbf{R}(\beta)\mathbf{n}_i$, $\mathbf{R}(\beta)$ being the rotation operator depending on the chirality angle.

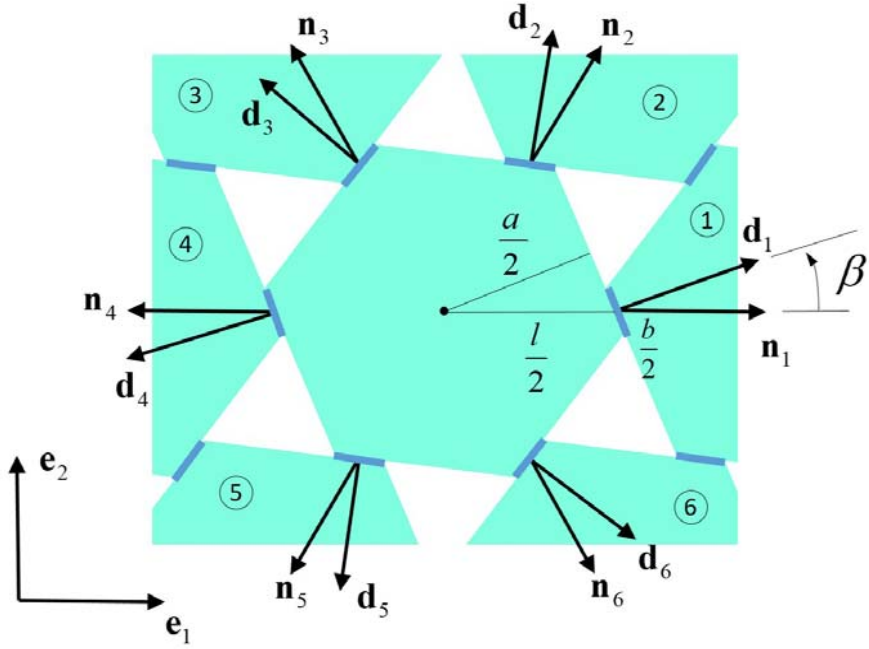

Fig. 3. The periodic hexagon and the unit vectors representing the material microstructure.

From the Cosserat homogenization presented in Section 2, the effective constitutive equations involving the elasticity tensors, given by equations (12) and (13), take the following matrix form

$$\begin{Bmatrix} \sigma_{11} \\ \sigma_{22} \\ \sigma_{12} \\ \sigma_{21} \\ m_1 \\ m_2 \end{Bmatrix} = \begin{bmatrix} 2\mu+\lambda & \lambda & -A & A & 0 & 0 \\ \lambda & 2\mu+\lambda & -A & A & 0 & 0 \\ -A & -A & \mu+\kappa & \mu-\kappa & 0 & 0 \\ A & A & \mu-\kappa & \mu+\kappa & 0 & 0 \\ 0 & 0 & 0 & 0 & S & 0 \\ 0 & 0 & 0 & 0 & 0 & S \end{bmatrix} \begin{Bmatrix} \gamma_{11} \\ \gamma_{22} \\ \gamma_{12} \\ \gamma_{21} \\ \chi_1 \\ \chi_2 \end{Bmatrix}, \qquad (15)$$

where the five effective elastic moduli are defined

$$\mu = \frac{\sqrt{3}}{4}\left(\bar{K}+\hat{K}\right) = \frac{\sqrt{3}}{4}\left(K_n + K_t\right),$$

$$\lambda = \frac{\sqrt{3}}{4}\left(\bar{K}-\hat{K}\right) = \frac{\sqrt{3}}{4}\left(K_n - K_t\right)\cos 2\beta,$$

$$\kappa = \frac{\sqrt{3}}{2}\left(K_n \sin^2 \beta + K_t \cos^2 \beta\right), \quad (16)$$

$$A = \frac{\sqrt{3}}{4}\left(K_n - K_t\right)\sin 2\beta,$$

$$S = \sqrt{3}\left[K_\varphi - \frac{a^2}{4\cos^2 \beta}\left(K_n \sin^2 \beta + K_t \cos^2 \beta\right)\right],$$

which depend on the constitutive parameters of the interfaces, on the cell size $a$ and on the chirality angle $\beta$. Here $\lambda$ and $\mu$ are the Lamé constants while $\kappa$, $A$ and $S$ are the elastic moduli associated to the micropolarity of the model, $A$ being the modulus that has an odd dependence on the chirality angle. Moreover, from this constitutive equation it easy found that if symmetric macro-fields are considered, the constitutive equations become those of a transversely isotropic material in a Cauchy continuum with in-plane elastic moduli:

$$E_{\text{hom}} = \frac{2\sqrt{3}K_n K_t \left(K_n + K_t\right)}{K_t \left(3K_n + K_t \cos^2 \beta\right) + K_n^2 \sin^2 \beta},$$

$$\nu_{\text{hom}} = \frac{\left(K_n - K_t\right)\left(K_t \cos^2 \beta - K_n \sin^2 \beta\right)}{K_t \left(3K_n + K_t \cos^2 \beta\right) + K_n^2 \sin^2 \beta}. \quad (17)$$

It is worth to note that the material may be auxetic for some values of the chirality angle $\beta$ and the ratio $K_t/K_n = k_t/k_n$. For instance, if the micro-beams interface shown in Appendix is applied, low values of the ratio $k_t/k_n$ may be obtained and the Poisson ratio takes negative values if $s/h < \tan \beta$. Moreover, a vanishing Poisson ratio may be obtained when $k_t = k_n$ or for chirality angle $\beta = \arctan \sqrt{k_t/k_n}$ (this condition is obtained in the micro-beam interface for $\beta = \arctan \sqrt{s/h}$). As shown in Figure 4, negative values may be obtained increasing the chirality angle, i.e. a relation between chirality and auxeticity is found. Moreover, for $\beta = 0$ the Cauchy elastic moduli by Bacigalupo and Gambarotta (2017) are recovered. This dependence of the Poisson ratio on the chirality is analogous with the results obtained on hexachiral lattices who exhibit an auxetic behavior (see for reference Prall and Lakes, 1997, Liu et al., 2012, and Bacigalupo and Gambarotta, 2014).

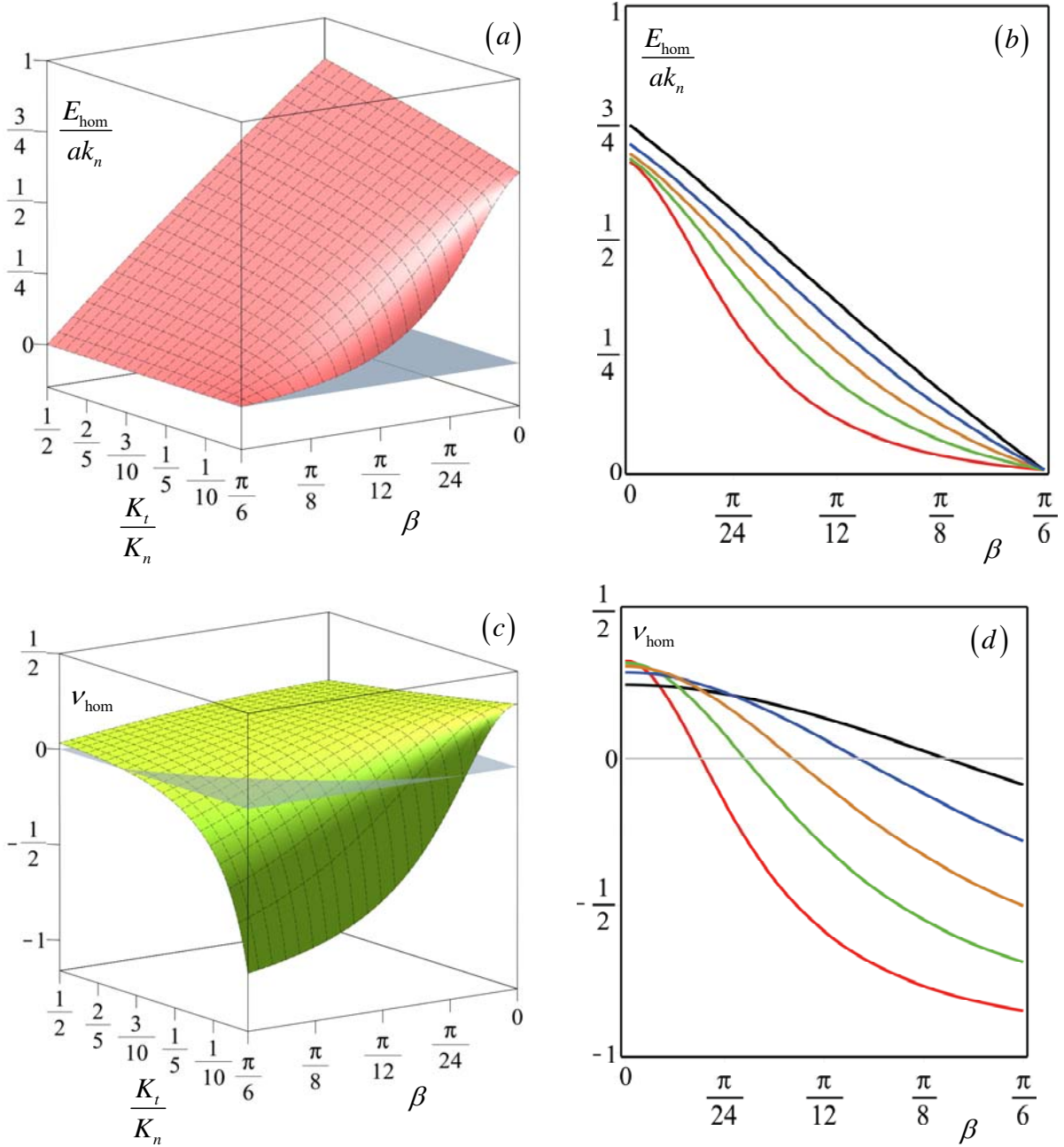

Figure 4. (a) Overall elastic modulus $E_{\text{hom}}$ in terms of chirality angle $\beta$ and stiffness ratio $K_t/K_n = k_t/k_n$; (b) $E_{\text{hom}}$ in terms of $\beta$ for different $k_t/k_n$; (c) Overall Poisson ratio $\nu_{\text{hom}}$ stiffness ratios $E_{\text{hom}}$ in terms of chirality angle $\beta$ and stiffness ratio $k_t/k_n$; (d) $\nu_{\text{hom}}$ in terms of $\beta$ for different $k_t/k_n$. In (b) and (d): black line $k_t/k_n = 1/5$, blue line $k_t/k_n = 1/10$, gold line $k_t/k_n = 1/20$, green line $k_t/k_n = 1/40$, red line $k_t/k_n = 1/100$.

In Figure 4, the dimensionless overall elastic modulus $E_{hom}/ak_n$ and the overall Poisson ratio $\nu_{hom}$ are shown in terms of the chirality angle $\beta$ and the interface stiffness ratio $K_t/K_n = k_t/k_n$. The overall elastic modulus decreases to increasing the chirality angle until it vanishes when $\beta = \pi/6$ and the interface length $b$ is equal to zero. A strong influence on the overall elastic modulus occurs for small values of the interface stiffness ratio $k_t/k_n$ (see Figures 4.a and 4.b). The diagrams of Figures 4.c and 4.d show a strong dependence of the overall Poisson ratio $\nu_{hom}$ on the chirality angle. In general, when increasing the chirality angle the auxecticity appears, and the overall Poisson ratio takes values increasingly negative with decreasing the interface stiffness ratio up to values close too -1. Such condition may be obtained through micro-structured interfaces, an example of which is given in the Appendix.

The wave propagation in the discrete blocky system is analysed by solving the eigen-problem in the equation (7), once specialized the components of the sub-matrices (8) of the hermitian matrix $\mathbf{C}_{Lag}$ that take the forms

$$\begin{aligned}
a_{11} &= \frac{1}{2}(4f_1 + f_2 + f_3)\bar{K} + \sqrt{3}(f_3 - f_2)\tilde{K} + \frac{3}{2}(f_2 + f_3)\hat{K}, \\
a_{22} &= \frac{3}{2}(f_2 + f_3)\bar{K} - \sqrt{3}(f_3 - f_2)\tilde{K} + \frac{1}{2}(4f_1 + f_2 + f_3)\hat{K}, \\
a_{12} &= a_{21} = \frac{\sqrt{3}}{2}(f_2 - f_3)\bar{K} + (2f_1 - f_2 - f_3)\tilde{K} + \frac{\sqrt{3}}{2}(f_3 - f_2)\hat{K}, \\
b_1 &= \frac{a}{2\cos\beta}\left[(2g_1 + g_2 - g_3)\tilde{K} - \sqrt{3}(g_2 + g_3)\hat{K}\right], \\
b_2 &= \frac{a}{2\cos\beta}\left[\sqrt{3}(g_2 + g_3)\tilde{K} + (2g_1 + g_2 - g_3)\hat{K}\right], \\
C &= 2K_\varphi(f_1 + f_2 + f_3) + \frac{a^2}{2\cos^2\beta}\hat{K}(6 - f_1 - f_2 - f_3),
\end{aligned} \quad (18)$$

being $f_i = f_i(\mathbf{k}, \mathbf{n}_i) = 1 - \cos(a\mathbf{k}\cdot\mathbf{n}_i/\cos\beta)$ and $g_i = g_i(\mathbf{k}, \mathbf{n}_i) = \sin(a\mathbf{k}\cdot\mathbf{n}_i/\cos\beta)$, $i$=1,6, and

$\hat{K} = \left(\sqrt{3}/3 - \tan\beta\right)\left(k_n \sin^2\beta + k_t \cos^2\beta\right)a$, $\qquad \bar{K} = \left(\sqrt{3}/3 - \tan\beta\right)\left(k_n \cos^2\beta + k_t \sin^2\beta\right)a$,

$\tilde{K} = a(k_n - k_t)\left(\sqrt{3}/3 - \tan\beta\right)\sin 2\beta / 2$.

The dispersive functions describing the acoustic properties of the equivalent micropolar continuum are obtained, once specialized the submatrices (14) with the constitutive equations (15) and (16), by solving the hermitian eigenvalue problem

$$\mathbf{H}_{\text{hom}}(\omega,\mathbf{k})\hat{\mathbf{V}} = \left[\mathbf{C}_{\text{hom}}(\mathbf{k}) - \omega^2\mathbf{M}\right]\hat{\mathbf{V}} = \mathbf{0}, \tag{19}$$

where the independent components of the hermitian matrix $\mathbf{H}_{\text{hom}}$ take the form

$$\begin{aligned}
H_{11}^{hom} &= (2\mu+\lambda)k_1^2 + (\mu+\kappa)k_2^2 - 2Ak_1k_2 - \rho\omega^2, \\
H_{22}^{hom} &= (\mu+\kappa)k_1^2 + (2\mu+\lambda)k_2^2 + 2Ak_1k_2 - \rho\omega^2, \\
H_{33}^{hom} &= S(k_1^2 + k_2^2) + 4\kappa - I\omega^2, \\
H_{12}^{hom} &= A(k_1^2 - k_2^2) + (\mu-\kappa+\lambda)k_1k_2, \\
H_{13}^{hom} &= -2i(\kappa k_2 - Ak_1), \\
H_{23}^{hom} &= -2i(\kappa k_1 + Ak_2).
\end{aligned} \tag{20}$$

The three dispersion functions depend on the chirality angle through equations (16). In case of achiral pattern $\beta = 0$ the modulus $A$ disappears, and the standard eigen-problem associated to plane propagation in a transversely isotropic micropolar continuum is obtained.

In Figure 5.a the dispersion surfaces are shown where the dimensionless frequency $\omega\sqrt{M/k_n}a$ is plotted in terms of the dimensionless wave vector components $k_i l$, $i=1,2$ for the ratio $k_t/k_n = 1/100$ and $\beta = \pi/8$. Two acoustic surfaces and an optical one result from the eigenproblem (19), which are periodic in the space of the wave vectors according to the hexagonal symmetry of the microstructure. A dense spectrum is obtained with Dirac cones, whose vertices are crossing points (like Dirac points) connecting the higher acoustic surface to the optical one, a common circumstance in hexagonal topologies (Lu *et al.*, 2013). A detail of the Dirac cones with Dirac points is shown in Figure 5.b. To appreciate the dependence of the dispersion curves on the chiral and mechanical parameters, the dispersion curves plotted along the edge of the triangular sub-domain of the first Brillouin zone (*reduced Brillouin zone*) are shown in the diagrams of Figures 5.c and 5.d. In Figure 5.c the frequency spectrum as function of the arch length $\Xi$ on the closed polygonal curve $\Upsilon$ with vertices identified by the values $\Xi_j$, $j = 0,..,3$, is plotted for $\beta = \pi/8$ and for two different ratios $k_t/k_n = 1/100$ (red line) and $k_t/k_n = 1/5$ (black line), respectively. It comes out that an increase of the ratio $k_t/k_n$ implies an homothetical increase of the frequency spectrum and hence of the Dirac point.

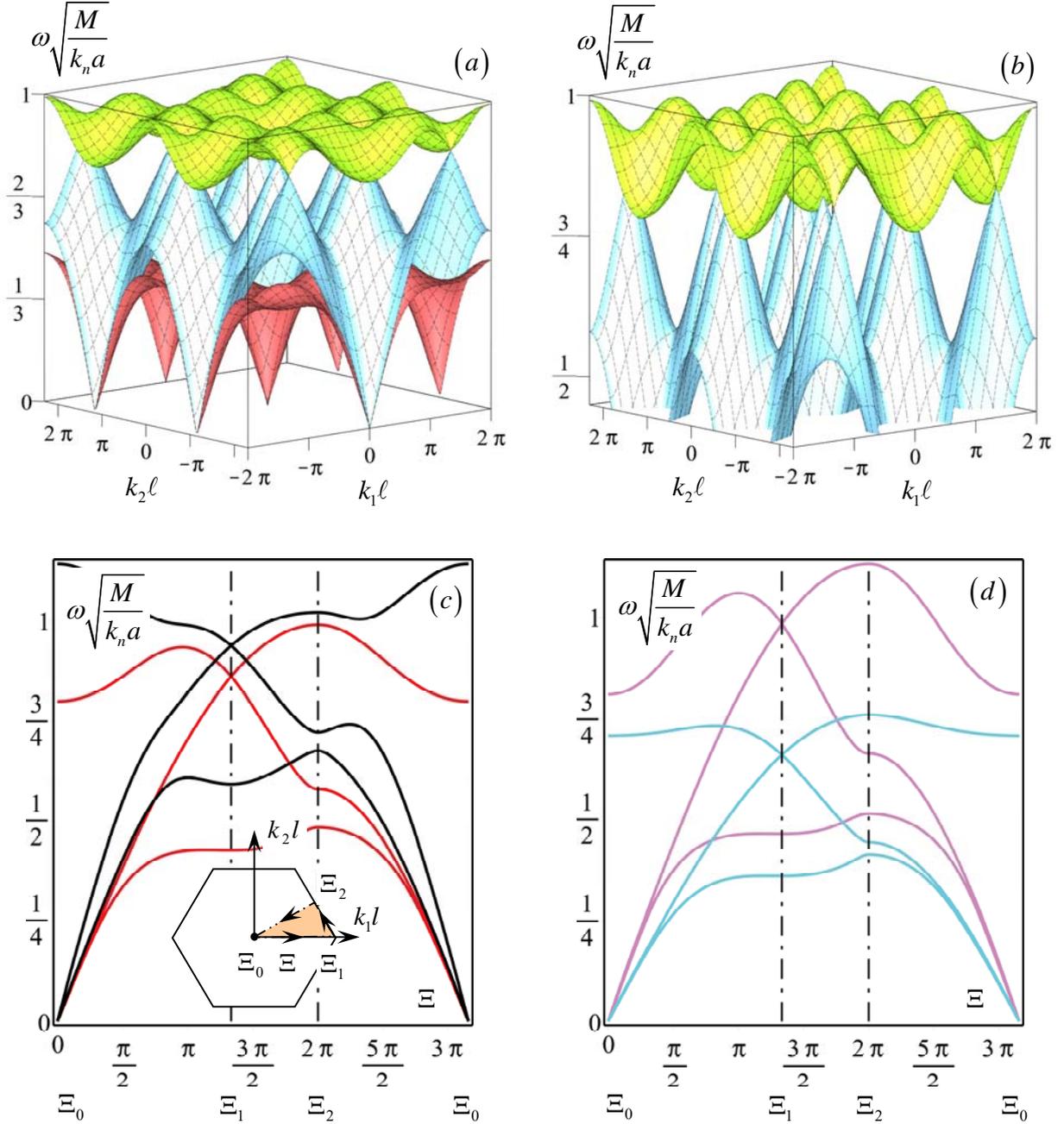

Figure 5. (a) Dispersion surface for $k_t/k_n = 1/100$ and $\beta = \pi/8$; (b) Detail of the Dirac cones; (c) Influence of the interface stiffness ratio $k_t/k_n$ for $\beta = \pi/8$ on the band structure along the closed polygonal curve $\Upsilon$ (red line $k_t/k_n = 100$, black line $k_t/k_n = 1/5$); (d) Influence of the chirality angle $\beta$ for stiffness ratio $k_t/k_n = 100$ (cyan line $\beta = \pi/7$, black line $\beta = \pi/9$).

In Figure 5.d the frequency spectrum is plotted for $k_t/k_n = 1/100$ and for two different chirality angle $\beta = \pi/7$ (cyan line) and $\beta = \pi/9$ (violet line), respectively. It comes out that an increase of the chiral angle $\beta$ implies an homothetical decrease of the frequency spectrum and hence of the Dirac point. Finally, the agreement between the dispersion functions obtained via continualized model in micropolar standard continuum is shown by comparing with the corresponding curves from the Lagrangian model in the homothetic subdomain of the reduced Brillouin zone bounded by the closed polygonal curve $\Upsilon$ for some values of the model parameters (see Figures 6.a and 6.b). In the considered cases a good accuracy of the micropolar model is achieved.

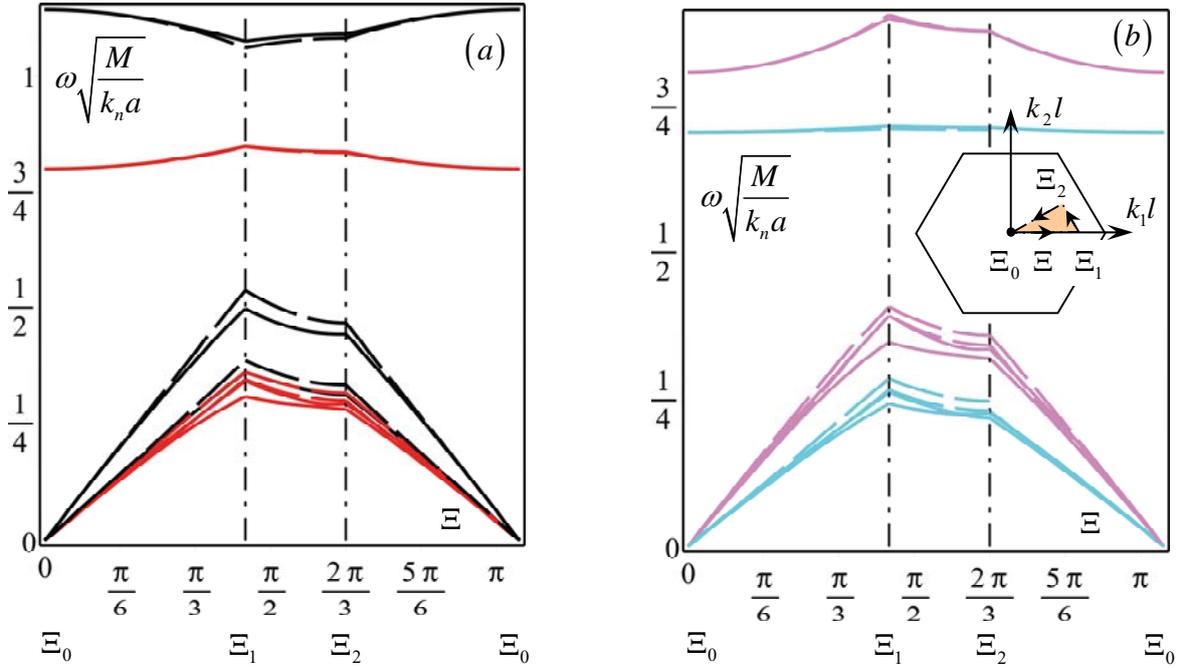

Figure 6. Comparison between the discrete model (continuous line) and the micropolar continuum model (dashed line) in the homothetic subdomain of the reduced Brillouin zone: (a) dispersion curves for $\beta = \pi/8$ and values of $k_t/k_n = 1/100$ (red line) and $k_t/k_n = 1/5$ (black line); (b) dispersion curves for $k_t/k_n = 1/100$ and values of $\beta = \pi/7$ (cyan line) and $\beta = \pi/9$ (violet line).

## 4. Tetrachiral blocky system

The tetrachiral system is made up of square blocks with side $a$ and inclined by the chiral angle $\beta$ ($0 \leq \beta < \pi/4$) as shown in Figure 4. The distance between the centres of the block is $l = a/\cos\beta$ and the interface length is $b = (1-\tan\beta)a$. The following relevant data are $A_b = a^2$, $A_{cell} = l^2 = a^2/\cos^2\beta$, $\rho_b = M/A_b$, $\rho = \rho_b \cos^2\beta$, $J = \rho_b a^4/6$, $r_b^2 = a^2/6$, $I = \rho a^2/6$. The overall elastic parameters of the interfaces are $K_n = K_n^i = k_n b = (1-\tan\beta)k_n a$, $K_t = K_t^i = k_t b = (1-\tan\beta)k_t a$ and $K_\varphi = K_\varphi^i = (1-\tan\beta)^3 a^3 k_n/12$, respectively. Also for this geometry of the microstructure the unit vectors identifying the blocks surrounding the reference one are independent on the chirality angle and are $\mathbf{n}_1 = -\mathbf{n}_3 = \mathbf{e}_1$, $\mathbf{t}_1 = -\mathbf{t}_3 = \mathbf{e}_2$, $\mathbf{n}_2 = -\mathbf{n}_4 = \mathbf{e}_2$, $\mathbf{t}_2 = -\mathbf{t}_4 = -\mathbf{e}_1$, while the unit vectors normal to the interfaces are $\mathbf{d}_i = \mathbf{R}(\beta)\mathbf{n}_i$, $\mathbf{R}(\beta)$ being the rotation operator depending on the chirality angle.

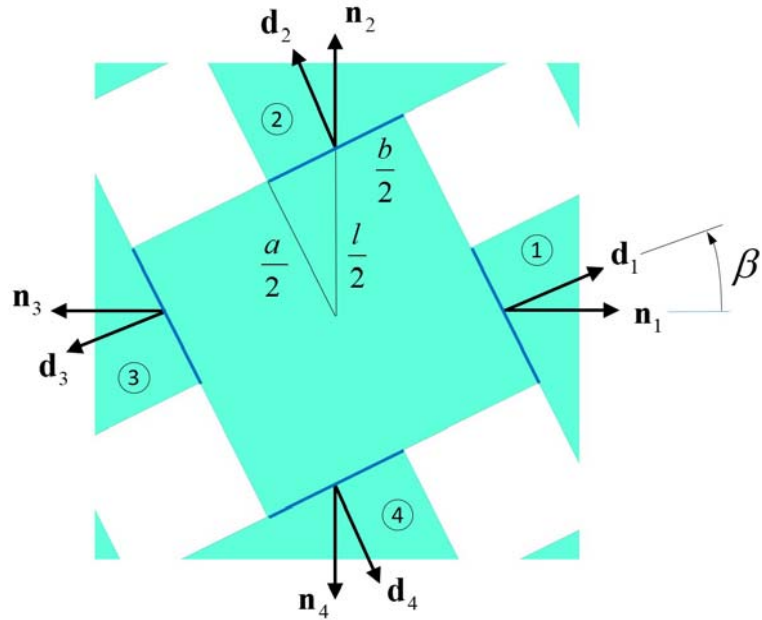

Fig. 7. The periodic square block and the unit vectors representative of the tetrachiral material structure.

The effective constitutive equations involving the elasticity tensors given by equations (12) and (13) take the following matrix form

$$\begin{Bmatrix} \sigma_{11} \\ \sigma_{22} \\ \sigma_{12} \\ \sigma_{21} \\ m_1 \\ m_2 \end{Bmatrix} = \begin{bmatrix} 2\mu & 0 & 0 & A & 0 & 0 \\ 0 & 2\mu & -A & 0 & 0 & 0 \\ 0 & -A & \kappa & 0 & 0 & 0 \\ A & 0 & 0 & \kappa & 0 & 0 \\ 0 & 0 & 0 & 0 & S & 0 \\ 0 & 0 & 0 & 0 & 0 & S \end{bmatrix} \begin{Bmatrix} \gamma_{11} \\ \gamma_{22} \\ \gamma_{12} \\ \gamma_{21} \\ \chi_1 \\ \chi_2 \end{Bmatrix}, \qquad (21)$$

with four effective elastic moduli

$$\begin{aligned}
\mu &= \frac{\overline{K}}{2} = \frac{1}{2}\left(K_n \cos^2 \beta + K_t \sin^2 \beta\right), \\
\kappa &= \hat{K} = K_n \sin^2 \beta + K_t \cos^2 \beta, \\
A &= \tilde{K} = \frac{\left(K_n - K_t\right)}{2}\sin 2\beta, \\
S &= \left(K_\varphi - \frac{1}{2}\frac{a^2}{\cos^2 \beta}\hat{K}\right) = \left[K_\varphi - \frac{1}{2}\frac{a^2}{\cos^2 \beta}\left(K_n \sin^2 \beta + K_t \cos^2 \beta\right)\right],
\end{aligned} \qquad (22)$$

depending on the constitutive parameters of the interface, on the cell size $a$ and on the chirality angle $\beta$. The square symmetry of the achiral system $(\beta = 0)$ implies a single Lamé constant $\mu$ and three elastic moduli associated to the micropolar character of the model, i.e. $\kappa$, $A$ and $S$. Also, for this system, the elastic constant $A$ become vanishing for achiral microstructure. In case of symmetric macro-fields the effective fourth order elasticity tensor $\mathbb{C}$ has the elasticities of the tetragonal system with symmetry rotations $\mathbf{R}_3^{\pi/2}$. In the Voigt notation, the constitutive equation is written in the form

$$\begin{Bmatrix} \Sigma_{11} \\ \Sigma_{22} \\ \Sigma_{12}^s \end{Bmatrix} = \begin{bmatrix} C_{1111} & C_{1122} & C_{1112} \\ C_{1122} & C_{1111} & -C_{1112} \\ C_{1112} & -C_{1112} & C_{1212} \end{bmatrix} \begin{Bmatrix} E_{11} \\ E_{22} \\ 2E_{12} \end{Bmatrix}, \qquad (23)$$

with

$$\begin{aligned}
C_{1111} &= C_{2222} = 2\mu - \frac{A^2}{2\kappa}, \\
C_{1122} &= -\frac{A^2}{2\kappa}, \quad C_{1212} = \frac{\kappa}{2}, \\
C_{1112} &= -C_{2212} = \frac{A}{2}.
\end{aligned} \qquad (24)$$

The elastic compliant matrix

$$\mathbf{D} = \frac{1}{2\kappa\mu - A^2} \begin{bmatrix} \kappa & 0 & -A \\ 0 & \kappa & A \\ -A & A & 4\mu \end{bmatrix}, \qquad (25)$$

is characterized by a vanishing overall Poisson coefficient and the coupling between the extensional strains and the shearing strains is governed by the chiral constant $A$.

A polar description of the overall elastic behavior is obtained by considering the uniaxial tension applied along the direction $\mathbf{n} = \cos\theta \mathbf{e}_1 + \sin\theta \mathbf{e}_2$ and evaluating the extension along $\mathbf{n}$ and the transverse extension. The resulting overall elastic modulus along direction $\mathbf{n}$ and Poisson ratio are expressed in the form

$$E_{\text{hom}}(\theta) = \frac{2\kappa\mu - A^2}{\kappa\cos^4\theta - 4A\sin\theta\cos^3\theta + 16\mu\sin^2\theta\cos^2\theta + 4A\sin^3\theta\cos\theta + \kappa\sin^4\theta},$$

$$\nu_{\text{hom}}(\theta) = \frac{\sin 2\theta \left[ 4A\cos 2\theta + (\kappa - 8\mu)\sin 2\theta \right]}{2\left(\kappa\cos^4\theta - 4A\sin\theta\cos^3\theta + 16\mu\sin^2\theta\cos^2\theta + 4A\sin^3\theta\cos\theta + \kappa\sin^4\theta\right)}. \qquad (26)$$

Both the non-dimensional overall elastic modulus $E_{\text{hom}}(\theta)/ak_n$ and the overall Poisson ratio $\nu_{\text{hom}}(\theta)$ are shown in the non-dimensional diagrams of figures 8 for different values of the chirality angle $\beta$ and of the interface stiffness ratio $k_t/k_n$. In Figures 8.a and 8.b the polar diagrams show the strong dependence of $E_{\text{hom}}(\theta)/ak_n$ on the angle $\theta$. This indicates a strong anisotropy with well-defined directions of maximum stiffness which depend on the angle of chirality $\beta$. In Figures 8.a the case $\beta = \pi/5$ is considered for different values of the ratio $k_t/k_n$ that does not affect the direction of maximum and minimum stiffness. As expected, increasing the interface stiffness ratio a higher overall elastic modulus is obtained. Conversely, in Figure 8.b the case $k_t/k_n = 1/100$ is considered for different values of the chirality angle $\beta$ that affects the direction of maximum and minimum stiffness. Here, increasing the chirality angle a decreasing of the overall elastic modulus is found. In Figures 8.c and 8.d the polar diagrams show the strong dependence of $\nu_{\text{hom}}(\theta)$ on the angle $\theta$. Specifically, in these diagrams the negative and positive values of the overall Poisson ratio are plotted in continuous and dashed line, respectively. It is worth to note that along the direction of minimum (maximum) overall elastic modulus, the overall Poisson ratio turns out to be positive (negative), so exhibiting a non-auxetic (auxetic) behaviour.

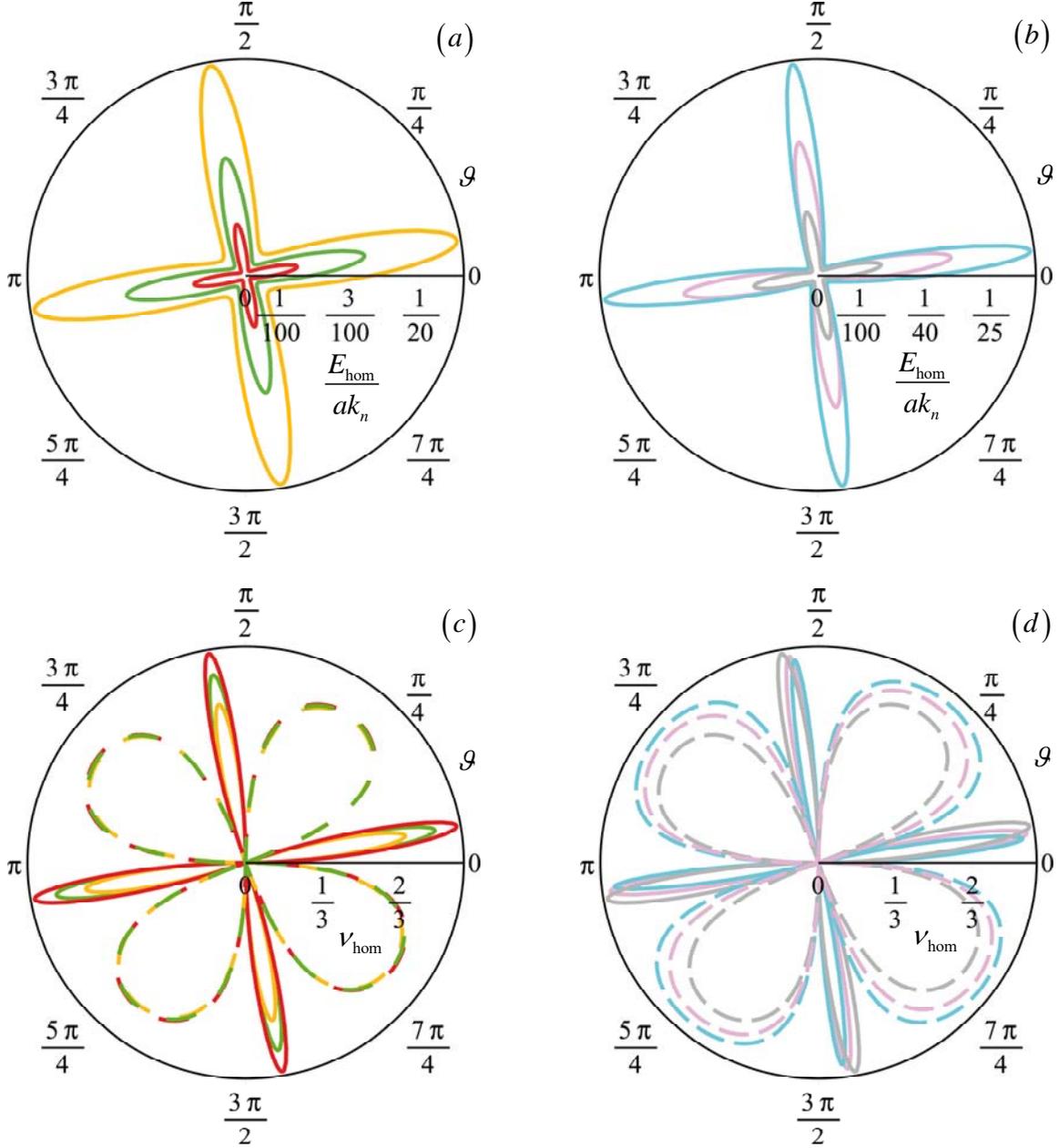

Figure 8. Polar diagrams of the overall elastic modulus $E_{\text{hom}}$ and Poisson ratio $\nu_{\text{hom}}$ for varying angular coordinate $\theta$. (a) $E_{\text{hom}}$ for $\beta = \pi/5$ and for different $k_t/k_n$; (b) $E_{\text{hom}}$ for $k_t/k_n = 1/100$ and for different $\beta$; (c) $\nu_{\text{hom}}$ for $\beta = \pi/5$ and for different $k_t/k_n$; (d) $\nu_{\text{hom}}$ for $k_t/k_n = 1/100$ and for different $\beta$. In (a) and (c) - red line $k_t/k_n = 1/100$, green line $k_t/k_n = 1/40$, gold line $k_t/k_n = 1/20$; in (b) and (d) - gray line $\beta = \pi/5$, pink line $\beta = \pi/6$, cyan line $\beta = \pi/7$; in (c) and (d) negative part in solid line and positive part in dashed line

Moreover, along direction $\theta = 0$ equation (26) provides $\nu_{hom} = 0$, namely a zero expansion behaviour. In Figures 8.c the case $\beta = \pi/5$ is considered for different values of the ratio $k_t/k_n$ that does not affect the direction of maximum and minimum overall Poisson ratio. It is worth to note that the interface stiffness ratio has a small influence on the positive values of $\nu_{hom}$, while some influence is observed on the negative values. Here, the auxetic behavior is marked for low values of the interface stiffness ratio. In Figure 8.d the case $k_t/k_n = 1/100$ is considered for different values of the chirality angle $\beta$ that affects the direction of maximum and minimum $\nu_{hom}$. Here, a weak influence of the chirality angle on the negative values of the overall Poisson ratio is observed. In Figure 9 the minimum overall Poisson ratio $\nu_{hom}^{min}$ is shown as a function of the ratio $K_t/K_n = k_t/k_n$ and the chirality angle $\beta$. In case of achiral geometry the minimum $\nu_{hom}$ vanishes, while increasing negative values are obtained when increasing both the chirality angle and the decreasing the interface stiffness ratio, up to values close to -1. Again, such condition may be obtained through micro-structured interfaces, an example of which is given in the Appendix.

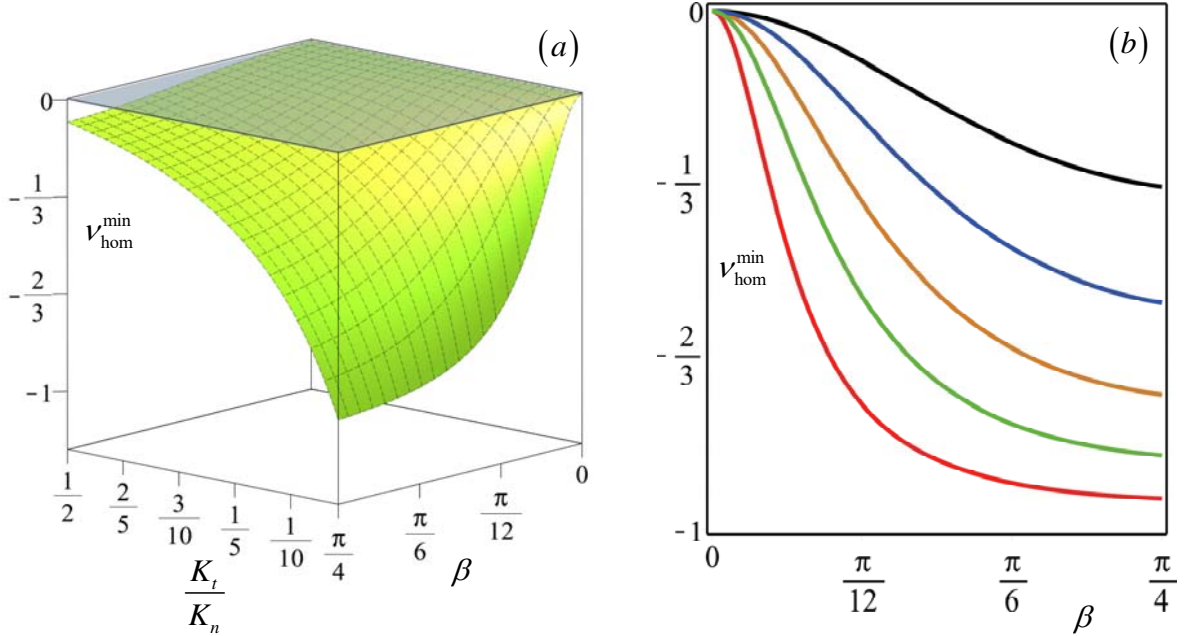

Figure 9. Minimum overall Poisson ratio $\nu_{hom}^{min}$ for varying chirality angle $\beta$ and stiffness ratio $k_t/k_n$ (black line $k_t/k_n = 1/5$, blue line $k_t/k_n = 1/10$, gold line $k_t/k_n = 1/20$, green line $k_t/k_n = 1/40$, red line $k_t/k_n = 1/100$).

The wave propagation in the discrete square blocky system is analysed by solving the eigen-problem (7), once specialized the components of the sub-matrices (8) of the hermitian matrix $\mathbf{C}_{Lag}$ as follows

$$\begin{aligned}
a_{11} &= 2\left(f_1 \bar{K} + f_2 \hat{K}\right), \\
a_{22} &= 2\left(f_1 \hat{K} + f_2 \bar{K}\right), \\
a_{12} &= a_{21} = 2\left(f_1 - f_2\right)\tilde{K}, \\
b_1 &= \frac{a}{\cos \beta}\left(g_1 \tilde{K} - g_2 \hat{K}\right), \\
b_2 &= \frac{a}{\cos \beta}\left(g_1 \hat{K} + g_2 \tilde{K}\right), \\
C &= 2K_\varphi \left(f_1 + f_2\right) + \frac{a^2}{2\cos^2 \beta}\hat{K}\left(4 - f_1 - f_2\right),
\end{aligned} \qquad (27)$$

where $\hat{K} = (1 - \tan\beta)\left(k_n \sin^2 \beta + k_t \cos^2 \beta\right)a$, $\bar{K} = (1 - \tan\beta)\left(k_n \cos^2 \beta + k_t \sin^2 \beta\right)a$, $\tilde{K} = a(k_n - k_t)(1 - \tan\beta)\sin 2\beta / 2$, $f_i = \left[1 - \cos(k_i a / \cos\beta)\right]$ and $g_i = \sin(k_i a / \cos\beta)$, $i=1,2$, respectively.

When considering the micropolar homogeneous effective continuum, the eigen-problem governing the propagation of harmonic plane waves is $\mathbf{H}_{\text{hom}}(\omega, \mathbf{k})\hat{\mathbf{V}} = \left[\mathbf{C}_{\text{hom}}(\mathbf{k}) - \omega^2 \mathbf{M}\right]\hat{\mathbf{V}} = \mathbf{0}$, where the independent components of the hermitian matrix $\mathbf{H}_{\text{hom}}$ take the form

$$\begin{aligned}
H_{11}^{\text{hom}} &= \left(2\mu k_1^2 + k_2^2 \kappa\right) - \rho\omega^2, \\
H_{22}^{\text{hom}} &= \left(2\mu k_2^2 + k_1^2 \kappa\right) - \rho\omega^2, \\
H_{33}^{\text{hom}} &= S\left(k_i^2 + k_i^2\right) + \kappa\left[2 + \frac{1}{4}\left(k_1^2 l^2 + k_2^2 l^2\right)\right] - I\omega^2, \\
H_{12}^{\text{hom}} &= A\left(k_1^2 - k_2^2\right), \\
H_{13}^{\text{hom}} &= i\left(k_1 A - k_2 \kappa\right), \\
H_{23}^{\text{hom}} &= i\left(k_1 \kappa - k_2 A\right).
\end{aligned} \qquad (28)$$

whose terms dependent on the chirality angle through equations (22). By solving problem (28) the dispersion functions and the corresponding wave vectors are obtained.

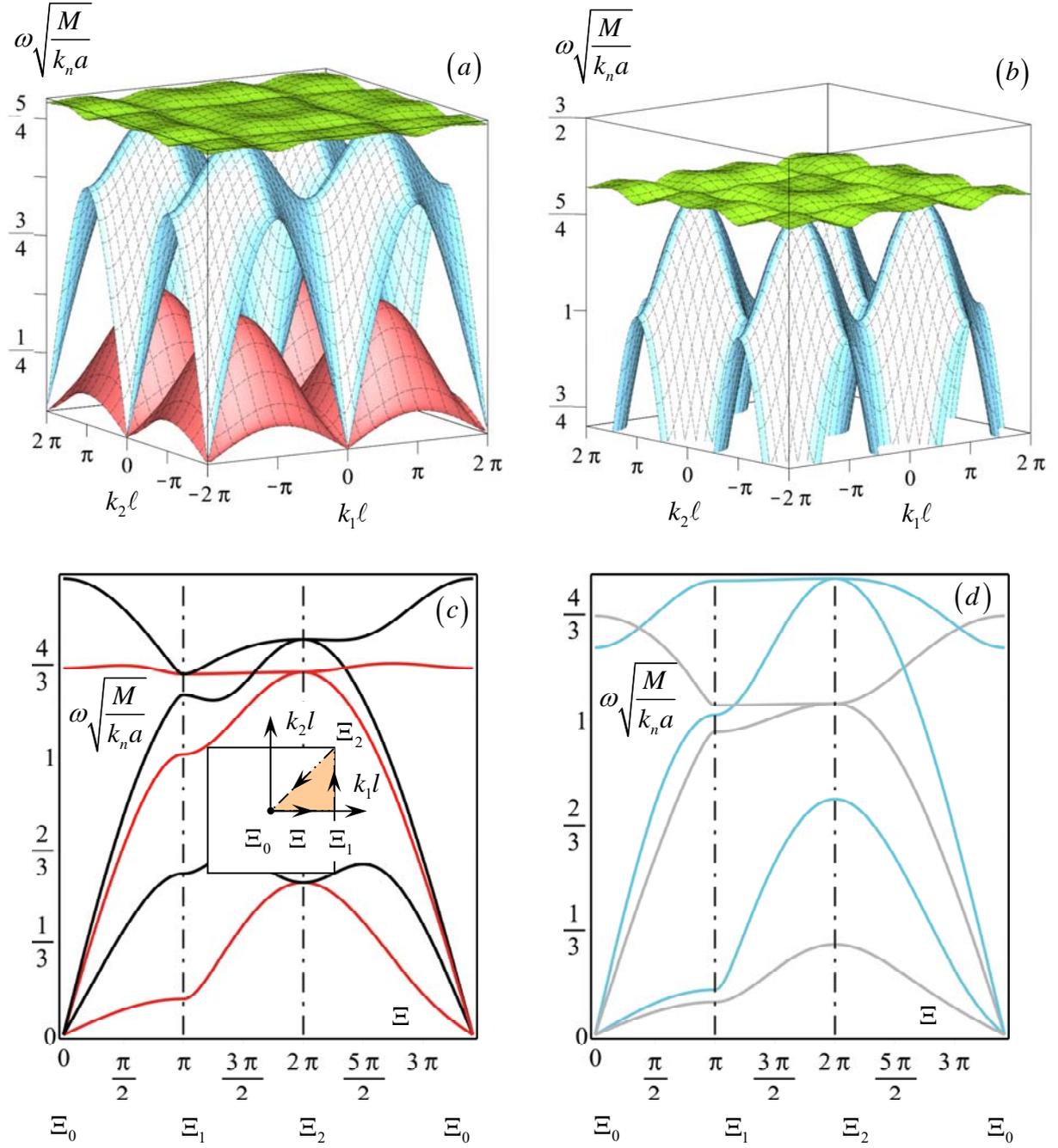

Figure 9. (a) Dispersion surface for $k_t/k_n = 1/100$ and $\beta = \pi/6$; (b) Detail of the quadratic point degeneracy; (c) Influence of the interface stiffness ratio $k_t/k_n$ for $\beta = \pi/6$ on the band structure along the closed polygonal curve $\Upsilon$ (red line $k_t/k_n = 100$, black line $k_t/k_n = 1/5$); (d) Influence of the chirality angle $\beta$ for stiffness ratio $k_t/k_n = 100$ (cyan line $\beta = \pi/7$, gray line $\beta = \pi/5$).

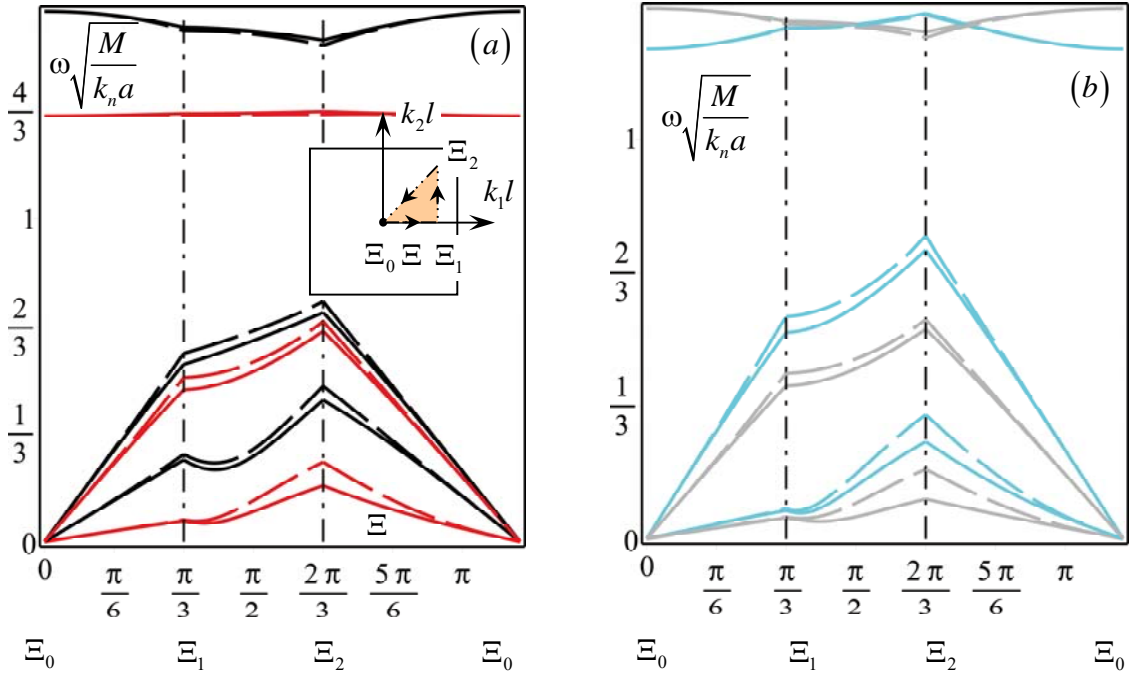

Figure 10. Comparison between the discrete model (continuous line) and the micropolar continuum model (dashed line) in the homothetic subdomain of the reduced Brillouin zone: (a) dispersion curves for $\beta = \pi/5$ and values of $k_t/k_n = 1/100$ (red line) and $k_t/k_n = 1/5$ (black line); (b) dispersion curves for $k_t/k_n = 1/100$ and values of $\beta = \pi/7$ (cyan line) and $\beta = \pi/5$ (gray line).

In Figure 9.a the dispersion surfaces are shown where the dimensionless frequency $\omega\sqrt{M/k_n a}$ is plotted in terms of the dimensionless wave vector components $k_i l$, $i=1,2$ for the ratio $k_t/k_n = 1/100$ and $\beta = \pi/6$. Two periodic acoustic surfaces and an optical one in the space of the wave vectors are recovered according to the two-dimensional cubic symmetry of the microstructure. A dense spectrum is obtained in which the upper acoustic surface and the optical one exhibit a band touching located at point $\Xi_2$, i.e. a quadratic point degeneracy (Lu *et al.*, 2013). A detail of this touching is shown in Figure 5.b. To appreciate the dependence of the dispersion curves on the chiral and mechanical parameters, the dispersion curves plotted along the edge of the triangular sub-domain of the first Brillouin zone (*reduced Brillouin zone*) are shown in the diagrams of Figures 9.c and 9.d. In Figure 9.c the frequency spectrum as function of the arch length $\Xi$ on the closed polygonal curve $\Upsilon$ with vertices identified by the values $\Xi_j$, $j = 0,..,3$, is plotted for $\beta = \pi/6$ and for two different ratios $k_t/k_n = 1/100$ (red line) and $k_t/k_n = 1/5$ (black line), respectively. It comes out that an increase of the ratio $k_t/k_n$ implies an homothetical increase of the frequency spectrum. In Figure 9.d the frequency spectrum is plotted for $k_t/k_n = 1/100$ and for two different chirality angle $\beta = \pi/7$ (cyan line) and $\beta = \pi/5$ (gray line), respectively. It comes out that an increase of the chiral angle $\beta$ implies an homothetical decrease of the frequency spectrum. Finally, the good agreement between the dispersion functions from the Lagrangian model and those from the micropolar standard continuum is shown in the homothetic subdomain of the reduced Brillouin zone bounded by the closed polygonal curve $\Upsilon$ for some values of the model parameters (see Figures 10.a and 10.b).

## 6. Conclusions

Two novel chiral block lattice topologies with centrosymmetric periodic cell have been proposed having interesting auxetic and acoustic behavior. The periodic material is obtained as a regular repetition of square or hexagonal rigid and heavy blocks connected by linear elastic interfaces, even in presence of micro-structured interfaces. To obtain the chiral configuration the blocks of the periodic pattern are equally rotated in the reference configuration and this rotation angle characterize the material chirality. The dynamic governing equation of the Lagrangian model have been analytically derived together with the constitutive equations of a micropolar equivalent continuum obtained in closed form via standard continualization. Moreover, the overall elastic moduli of the equivalent Cauchy continuum have been derived via proper condensation procedure from the micropolar constitutive equation. The frequency band structure of the Lagrangian model has been derived by solving a hermitian eigenvalue problem. An approximation of the Floquet-Bloch spectrum has been obtained through the equivalent micropolar model. This outcome is based on an eigenproblem governed by a hermitian matrix which approximates the corresponding one of the Lagrangian model with a second order Taylor polynomial in the wave vector variable.

The hexachiral blocky material has been described through five micropolar constitutive elastic moduli, of which four are those of the standard two-dimensional isotropic micropolar continuum and the fifth, coupling the angular (extensional) strains to the normal (shearing) stresses, is an odd function of the chirality angle. The corresponding elasticity constants of the equivalent isotropic Cauchy continuum, namely the overall elastic modulus and the overall Poisson ratio, have been obtained in terms of the normal and tangential stiffness of the interface, of the characteristic size of the block and of the chirality angle. It has been shown that through a tuning of the chirality angle and of the ratio between the tangent and normal stiffness of the interface it is possible to obtain different elastic behaviors, from strongly auxeticity to zero expansion response up to not-auxeticity. Moreover, the resulting acoustic properties show a high-density frequency spectrum in which Dirac cones are detected between the higher acoustic surface and the optical one. When analyzing the micropolar equivalent continuum, all the three branches in the frequency band structure turn out to be particularly accurate for wavelengths greater than 3-4 times the characteristic dimension of the hexagons.

The tetrachiral blocky material is described through four micropolar constitutive elastic moduli, of which three are those of the standard two-dimensional cubic micropolar continuum and the fourth is a coupling parameter odd function of the chirality angle in analogy to the hexachiral system. Due to the cubic material symmetry, the equivalent Cauchy continuum has three elastic moduli. Hence, the overall elastic modulus and the overall Poisson ratio, represented in a polar description along the direction of applied stress, present a cubic symmetry. While the overall elastic modulus strongly depends on both the chirality angle and the interface stiffness ratio, on the contrary the negative part of the overall Poisson ratio mainly depends on the interface stiffness ratio and its positive part mainly depends on the chirality angle. Although the material response is not isotropic, the minimum value of the overall Poisson ratio may approach values close to -1 when decreasing the interface stiffness ratio and increasing the chirality angle. The acoustic properties of the tetrachiral material present a high-density frequency spectrum in which the upper acoustic surface and the optical one exhibit a band touching consisting in a quadratic point degeneracy. Finally, also in this case, all the three branches in the frequency band structure obtained by the micropolar equivalent model result in good agreement with the corresponding ones of the Lagrangian model.

**Acknowledgements**

The authors acknowledge financial support of the (MURST) Italian Department for University and Scientific and Technological Research in the framework of the research MIUR Prin15 project 2015LYYXA8, Multi-scale mechanical models for the design and optimization of micro-structured smart materials and metamaterials, coordinated by prof. A. Corigliano. AB thankfully acknowledge financial support by National Group of Mathematical Physics (GNFM-INdAM).

**Appendix – Micro-structured interface**

Let us consider the simple interface made up of parallel beams shown in Figure A.1. Each beam is assumed to be elastic with Young modulus $E$, thickness $s$, height $h$ and unit depth. The distance between two adjacent beams is $d$.

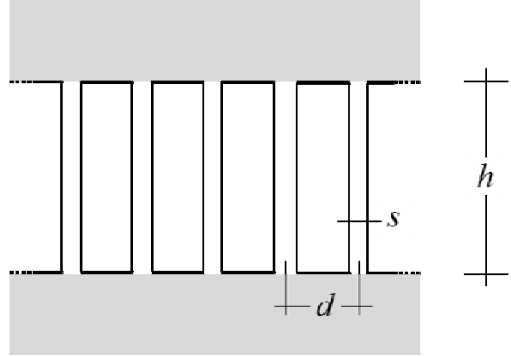

Figure A.1 The micro-structured interface with parallel beams.

Under the simple assumption of the Euler-Bernoulli model, the normal and tangential stiffness per unit length of the interface are

$$k_n = \frac{Es}{hd} \quad , \quad k_t = \frac{E}{d}\left(\frac{s}{h}\right)^3 \tag{29}$$

and the ratio is

$$\frac{k_t}{k_n} = \left(\frac{s}{h}\right)^2 \tag{30}$$

that turns out to be independent on the beam distance $d$. When assuming values of the geometric parameters within the validity limits of the beam theory, i.e. $s/h \cong 1/20 \div 1/8$, very low values of the ratio $k_t/k_n$ are obtained, while reducing the distance $d$ it is possible to guarantee acceptable values of the axial stiffness of the interface.

In case of hexagonal chiral block assemblies, form equation (17) the overall elastic modulus and the overall Poisson ratio take the following form

$$E_{hom} = \frac{2\sqrt{3}\left(\frac{\sqrt{3}}{3} - tg\beta\right)\left(\frac{s}{h}\right)^3\left[1+\left(\frac{s}{h}\right)^2\right]\frac{a}{d}E}{\left[\left(\frac{s}{h}\right)^2\left(3+\left(\frac{s}{h}\right)^2\cos^2\beta\right)+\sin^2\beta\right]},$$

$$\nu_{hom} = \frac{\left[1-\left(\frac{s}{h}\right)^2\right]\left[\left(\frac{s}{h}\right)^2\cos^2\beta - \sin^2\beta\right]}{\left(\frac{s}{h}\right)^2\left[3+\left(\frac{s}{h}\right)^2\cos^2\beta\right]+\sin^2\beta}.$$

(31)

In Figure A.2 these quantities $E_{hom}d/Ea$ and $\nu_{hom}$ are shown in terms of the geometrical ratio $s/h$ and the chirality angle $\beta$.

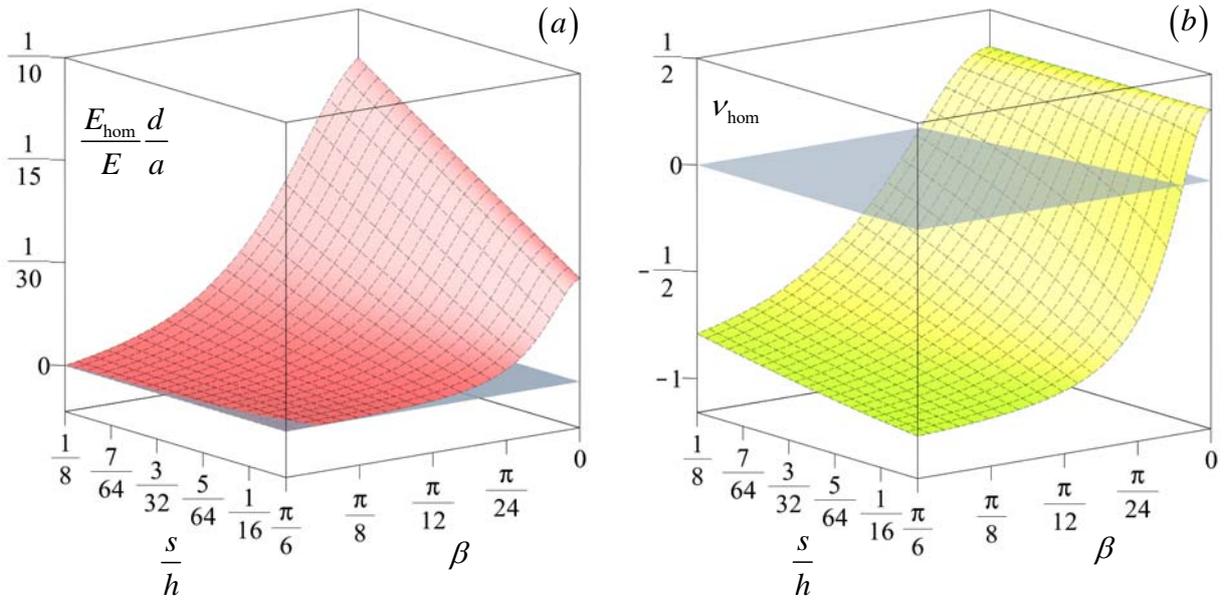

Figure A.2. (a) Non-dimensional overall elastic modulus; (b) Overall Poisson ratio.